\documentclass[twocolumn,showpacs,preprintnumbers,amsmath,amssymb]{revtex4}
\usepackage{graphicx}
\usepackage{dcolumn}
\usepackage{bm}
\usepackage{txfonts}

\begin{document}

\title{Relation among concentrations of incorporated Mn atoms, ionized Mn acceptors\\
and holes in \textit{$p$}-(Ga,Mn)As epilayers}

\author{R. Moriya}
\author{H. Munekata}
\affiliation{Imaging Science and Engineering Laboratory\\
Tokyo Institute of Technology\\
4259 Nagatsuta, Midori-ku, Yokohama 226-8503, JAPAN\\
}
\date{\today}

\begin{abstract}

The amount of ionized Mn acceptors in various \textit{$p$}-type Mn-doped GaAs epilayers 
grown at high (\textit{$T_s$}  = 580\\ ‹C) and low (\textit{$T_s$} = 250 ‹C)
substrate has been evaluated by electrochemical capacitance-voltage measurements, and has been
compared systematically with concentrations of incorporated Mn atoms and holes for wide range
of Mn concentration (\textit{N$_{Mn}$} = 10$^{17}$ $\sim$  10$^{21}$ cm$^{-3}$). Low \textit{$T_s$}
samples are found to be classified into three different regions: (a) highly resistive, strongly
compensated region in \textit{N$_{Mn}$} $\leq$ 10$^{18}$ cm$^{-3}$, (b) low resistive, fully ionized region
in $N_{Mn}$ in between $5 \times 10^{18}$ and 10$^{20}$ cm$^{-3}$ in which impurity conduction
dominates, and (c) magnetic, low-resistive region in $N_{Mn}\geq 2\times 10^{20}$ cm$^{-3}$
in which magnetism and carrier transport are strongly correlated. Quantitative assessment
of anomalous Hall effect at room temperature is also carried out for the first time.
\end{abstract}


\maketitle

\section{Introduction}

(In,Mn)As \cite{1,2} and (Ga,Mn)As \cite{3,4}, are III-V-based magnetic alloy semiconductors (III-V-MAS)
that contain a large amount of magnetic ions. They are usually prepared by molecular beam epitaxy
(MBE) at substrate temperatures of $200 - 300$ ‹C, which results in suppression of second phases
under the condition beyond the equilibrium solubility limit of Mn ($\sim$ 10$^{19}$ cm$^{-3}$). They exhibit hole-induced
ferromagnetism, and because of this fact, they are expected as one of the strong candidate materials for 
spintronics devices. Experimental studies on these semiconductors \cite{5,6} have revealed that they could be 
described qualitatively as a random alloy in which Mn substitutes for a group III site and takes a dual role
of acceptor and local magnetic moment.

Doping of Mn in GaAs up to the equilibrium solubility limit has been
studied in 70's. Samples of GaAs:Mn were grown at relatively high temperatures ($\geq$ 560 ‹C) either by MBE or
boat-grown method. We call these samples as conventional GaAs:Mn in this paper. In the dilute limit, the level
of Mn acceptor was established to be \textit{E$_A$} $\approx$ 110 meV \cite{7}. Hole concentration could be
obtained from low-field Hall effect, through which reduced \textit{E$_A$} (\textit{E$_A$} = $60 - 80$ meV) was found
with increasing Mn concentration \textit{N$_{Mn}$} up to around 10$^{19}$ cm$^{-3}$ \cite{8}. Beyond 10$^{19}$
cm$^{-3}$, metallic impurity-band conduction took place at low temperatures \cite{9}. Simultaneously, Hall
effect deviated from the normal behavior and diminished at low temperatures. Filamental transport characteristics
were also inferred. On the other hand, as for the \textit{p}-type III-V-MAS (Ga,Mn)As grown at relatively low temperatures,
ferromagnetic order developed for \textit{N$_{Mn}$} $>$ 4 $\times$ 10$^{20}$ cm$^{-3}$. Metallic conduction was
found in wide range of temperatures for Mn concentrations \textit{N$_{Mn}$} ranging \textit{N$_{Mn}$} =$7 - 12$ $\times $
10$^{20}$ cm$^{-3}$ (0.03 $\leq$ x $\leq $ 0.05 in Ga$_{1-\textit{x}}$Mn$_\textit{x}$As) \cite{10}, whereas, beyond
\textit{N$_{Mn}$} = 12 $\times$ 10$^{20}$ cm$^{-3}$, non-metallic conduction took place. Magneto-transport was strongly
influenced by the magnetic behavior of samples, as exemplified by the anomalous Hall effect \cite{5}.

These comparisons raise a question whether (Ga,Mn)As can be viewed as the extension of conventional Mn doped \textit{p}-type
samples. In order to find answers to this question, development of physical methods that allow us to systematically 
estimate the concentrations of ionized Mn acceptors \textit{N$_{Mn^-}$} and holes \textit{p} for wide range of Mn concentrations is desired. 
In particular, determination of \textit{N$_{Mn^-}$} and \textit{p} is important for ferromagnetic samples, since the Curie temperature \textit{T$_c$} 
depends strongly on these quantities \cite{11}.

In this work, we are concerned with determination of \textit{N$_{Mn}$}, \textit{N$_{Mn^-}$}, and \textit{p}
in Mn-doped GaAs epitaxial layers with wide range of Mn concentrations (10$^{17} - 10^{21}$ cm$^{-3}$), deposited at low
(\textit{T$_s$} = 250 ‹C) and high (\textit{T$_s$} = 580 ‹C) substrate
temperatures by molecular beam epitaxy. $N_{Mn}$ and $p$ are determined by x-ray diffraction method and low-field Hall
effect, respectively. Key experiment is electrochemical $C - V$ method \cite{12}. This method appears to suppress the
tunneling current across the thin depletion region in heavily doped samples, and allow us to obtain the amount
of Mn incorporated in the form of ionized acceptor without the application of a magnetic field. Carefully comparing
\textit{N$_{Mn^-}$} with \textit{N$_{Mn}$} and \textit{p} leads us to the following conclusions. For high Ts samples,
Mn starts to precipitate as MnAs when \textit{N$_{Mn}$} exceeds beyond about 3 $\times$ 10$^{18}$ cm$^{-3}$. Except this fact,
electronic behaviors of epilayers prepared at high \textit{T$_s$} are essentially the same as those studied in 70's.
For low Ts samples, MnAs precipitates are not found for all the samples studied in this work
(\textit{N$_{Mn}$} $\leq$ 1.5 $\times$ 10$^{21}$ cm$^{-3}$). Electronic behaviors are found to be classified into 
three different regions. Firstly, (a) highly resistive, strongly compensated region in \textit{N$_{Mn}$} $\leq$
10$^{18}$ cm$^{-3}$, and secondly, (b) low-resistive, fully ionized region in \textit{N$_{Mn}$} = 
$5 \times 10^{18} - 10^{20}$ cm$^{-3}$ in which Hall effect virtually disappears. 
This peculiar behavior is reminiscent of the impurity conduction found in the conventional GaAs:Mn with \textit{N$_{Mn}$}
$\geq$ mid 10$^{18}$ cm$^{-3}$. In terms of ionization of incorporated Mn, this region shows \textit{N$_{Mn}$}
$\approx$ \textit{N$_{Mn^-}$}. The third region is (c) magnetic,
low-resistive region in \textit{N$_{Mn}$} $\geq$ 2 $\times$ 10$^{20}$ cm$^{-3}$ in which magnetism and carrier
transport are strongly correlated. Revival of Hall effect in this region suggests the occurrence of electrical
conduction through the valence band. In terms of ionization of Mn, compensation starts to take place again with
increasing \textit{N$_{Mn}$}. From the relation between \textit{N$_{Mn^-}$} and \textit{p} 
in the third region, we quantitatively assess the contribution of anomalous Hall effect at room temperature for the 
first time for magnetic (Ga,Mn)As.

\section{Experimental}

Epitaxial layers were grown on semi-insulating GaAs (100) substrates by molecular beam epitaxy.
In a growth chamber, a substrate surface was first thermally cleaned at the substrate temperature of \textit{T$_s$}
= 600 ‹C under the As$_4$ beam flux. Then, a 300-nm-thick GaAs buffer layer was grown at \textit{T$_s$} = 580 ‹C.
This was followed by the growth of either Mn- or Be-doped epitaxial layers. Three different types of epitaxial
layers were grown. The first type was Mn-doped layers grown at relatively low \textit{T$_s$} of 250 ‹C.
The second and third types were Mn- and Be-doped layers grown at \textit{T$_s$} = 580 ‹C, respectively.
The range of Mn and Be concentrations studied in this work were 2 $\times 10^{17} - 2 \times 10^{21}$
cm$^{-3}$ and 9 $\times 10^{17} - 1 \times 10^{20}$ cm$^{-3}$, respectively.
The ratio of As$_4$/Ga beam equivalent pressure (BEP) was kept to be 4 - 6 during the epitaxial growth.
The growth rate was typically 0.7 $\mu$m/hr, and in any samples, the thickness of the doped layer was kept to be
around 0.5 $\mu$m.

Mn concentration \textit{N$_{Mn}$} of epilayers with \textit{N$_{Mn}$} $\geq$ 1 $\times$10$^{20}$ cm$^{-3}$ was
determined on the basis of the relationship between strain-free cubic lattice constant \textit{a$_0$} and \textit{N$_{Mn}$}
described in ref.2. A perpendicular lattice constant \textit{a$_\perp$} was measured by a double-crystal x-ray diffractometer,
which was then converted into \textit{a$_0$} assuming that the epilayers were fully compressibly strained.
Taking into account the fact that a depends to some extend on the amount of excess As in the epitaxial layers \cite{13},
we set the margin of $\pm$ 0.5 $\times$ 10$^{20}$ cm$^{-3}$ for the estimation of \textit{N$_{Mn}$}.
This margin is also shown as a horizontal bar in the Figs.~\ref{NandN-}, ~\ref{Nandp} and ~\ref{asymmetric}. As to the relatively low \textit{N$_{Mn}$} region
(\textit{N$_{Mn}$} $<$ 10$^{20}$ cm$^{-3}$), an equilibrium vapor pressure curve of Mn \cite{14} was used to extrapolate
\textit{N$_{Mn}$} values from high concentration region. In this case, slight fluctuation of growth rate from MBE run to
run was also taken into account for estimating the margin of incorporated Mn concentration. Mn incorporation ratio was
assumed to be unity.

Determination of concentration of ionized Mn acceptors $N_{A^-}$ at room temperature was carried out by
electrochemical capacitance-voltage (ECV) method. This method, being routinely used in semiconductor laboratories,
measures the differential capacitance $C = dQ/dV$ of space charges in semiconductors and related interface \cite{15}.
Being analogous to the Schottkey junction at the metal-semiconductor interface, the difference in the chemical potential
of an electron between an electrolyte solution (ES) and a semiconductor gives rise to charge transfer across the
ES-semiconductor interface. This results in the formation of space charge region, being called the depletion region,
composed of ionized impurities in a semiconductor. When the amount of interface states is negligibly small and the
distribution of ionized impurities is uniform, the capacitance of depletion region for \textit{p}-type semiconductors
can be express in the following equation: 

\begin{equation}
\label{capacitance}
C=\sqrt{\frac{q\epsilon_s(N_{A^-}-N_{D^+)}}{V_{bi}-V-\frac{kT}{q}}}
\end{equation} 

Here, $q$, $\epsilon_s$, $(N_{A^-}-N_{D^+})$, $V_{bi}$, $V$, $kT$ are electron charge,
relative dielectric constant, the amount of uncompensated charges, built-in barrier height, bias voltage to a
semiconductor, and thermal energy, respectively. Impurities whose states $E_A$ are sufficiently less than $V_{bi}$
($V_{bi}$ = $0.5 - 0.8$ eV in $p$-GaAs) are fully ionized in the depletion region. This condition holds for Mn atoms
that are incorporated substitutionally, since $E_A$ is 110 meV or less in the doping range of 10$^{15} - 10^{19}$
cm$^{-3}$ \cite{7, 8} and satisfies the condition $E_A$ $<$ $V_{bi}$. Therefore, by plotting $1/C^2$ vs $V$
(the Schottkey plot), we are able to obtain the concentration of uncompensated ionized Mn acceptors, $N_{A^-} - N_{D^+}$,
together with $V_{bi}$ from the slope and the intersection to the $x$ axis, respectively \cite{16}.

Measurements were carried out by using a standard electrochemical cell equipped with a saturated calomel reference
electrode. $p$-type surface was placed on an o-ring through which electrical contact with the electrolyte solution of
0.1M C$_6$H$_2$(OH)$_2$(SO$_3$Na)$_2$¥H$_2$O was established. The contact area was approximately 0.01 cm$^2$.
A counter ohmic contact to the $p$-type semiconductor was achieved through soldered indium metal on the surrounding
area of the o-ring contact. The frequency $\omega$ used in our measurements was varied between 1 and 25 kHz.
Frequency dependence on capacitance was fairly small for high $T_s$ samples, whereas it appeared to be not negligibly
small for low $T_s$ samples. The $C - V$ region which shows only weak w dependence was used for data analysis in the
low $T_s$ samples. This point will be discussed in the later section. Depth profile, as carried out by combining
electrochemical etching and $dQ/dV$ measurement, showed that all the samples studied in this work have uniform
doping profile except for a few-nm-thick top surface region where a native oxide layer exists. Results shown in this
paper are based on the data measured after etching samples for the thickness of about 50 nm.

Hole concentration $p$ is estimated from low-field ($H$ = 0.3 Tesla) Hall effect measurements at room and
liquid-nitrogen temperatures. Van der Pauw method \cite{17} was used for 3-mm square samples. Ohmic contacts were
formed by soldering indium at each corner of the sample. In general, Hall resistance $R_H$ for samples incorporating
magnetic impurities can be expressed as,
\begin{equation}
\label{RH}
R_H=(R_\textrm{0}/d)B+(R_s/d)M
\end{equation}
\begin{subequations}
\begin{eqnarray}
\label{R0}
R_\textrm{0}=\gamma/ep\\
\notag\\
\label{Rs}
R_s=c\rho^n
\end{eqnarray}
\end{subequations}
, where $R_0$ and $R_s$ are ordinary and anomalous Hall coefficients, and $B$, $M$, $d$, $\rho$, and $c$ are
a magnetic field, magnetization, sample thickness, sample resistivity and the constant associated with asymmetric
scattering due to magnetic impurity \cite{18}. In the eq. (~\ref{R0}), the scattering factor $\gamma$ was assumed to be
unity. The power index $n$ in eq. (~\ref{Rs}) is either $n$ = 1 or 2 for skew or side-jump scattering, respectively.
For conventional GaAs:Mn with $N_{Mn}$ $\leq$ mid 10$^{18}$ cm$^{-3}$, $p$ was extracted directly from 
$R_H=(R_0/d)$, assuming that magnetization was negligibly small. For magnetic (Ga,Mn)As, however,
the contribution of the second term in eq. (~\ref{RH}) has to be taken into account \cite{3,5}. It has been demonstrated
that the second term can be suppressed when samples are at low temperature (50 mK) with a very high magnetic
field (27 Tesla) \cite{19}. This condition, however, is not available routinely in the semiconductor laboratories.
Furthermore, it can not be used for highly resistive and/or paramagnetic samples.

In this study, the contribution of the second term in eq. (~\ref{RH}) is evaluated by examining the relation among
$N_{Mn}$, $N_{Mn^-}$, and $p$ in two ways. The first approach, being applicable for $N_{Mn}$ below the
metal-insulating transition, is to examine the consistency among those quantities in view of charge neutrality
condition $n + N_{Mn^-} = p$ governed by the Fermi distribution \cite{20}. Experimental values are substituted
for $N_{Mn}$, $N_{Mn^-}$, and $p$ in eq. (~\ref{4a}) to extract $E_A$ and $E_F$. If there is mere contribution of the
second term in eq. (~\ref{RH}), it can be recognized as unreasonable $E_A$ and $E_F$ values that show deviation from those
values established by earlier works. The results of this approach are discussed in the section \ref{3-2}.
\begin{widetext}
\begin{subequations}
\begin{eqnarray}
\label{4a}
4\pi(2m_e/h^2)^{\frac{3}{2}}F_{\frac{1}{2}}(E_F)+\frac{N_A}{1+4exp(\frac{E_A-E_g-E_F}{kT})}=4\pi(2m_h/h^2)^{\frac{3}{2}}F_{\frac{1}{2}}[-(E_F+E_g)]\\
\notag\\
\label{4b}
F_n(\zeta)=\int_0^\infty\frac{E^ndx}{exp[\frac{E-\zeta}{kT}]+1},
\end{eqnarray}
\end{subequations}
\end{widetext}

The second approach is used for high impurity concentration that is beyond the metal-insulating transition.
In this case, impurities are fully ionized and the amount of current carriers is equal to the number of ionized
impurities \cite{21}. If there is no contribution of the second term in eq. (~\ref{RH}), the $p$ value obtained from the Hall
effect measurement should coincide with $N_{Mn^-}$ obtained from $C - V$ method. On the other hand, if we observe the
difference between the measured $N_{Mn^-}$ and $p$ values, it can be regarded as the contribution from the second term
in eq. (~\ref{RH}) and thus the asymmetric scattering. The results of this approach are discussed in the section \ref{3-3}.

\section{Results and discussion}
	\subsection{$I - V$ and $C - V$ characteristics}

\begin{figure}
\includegraphics{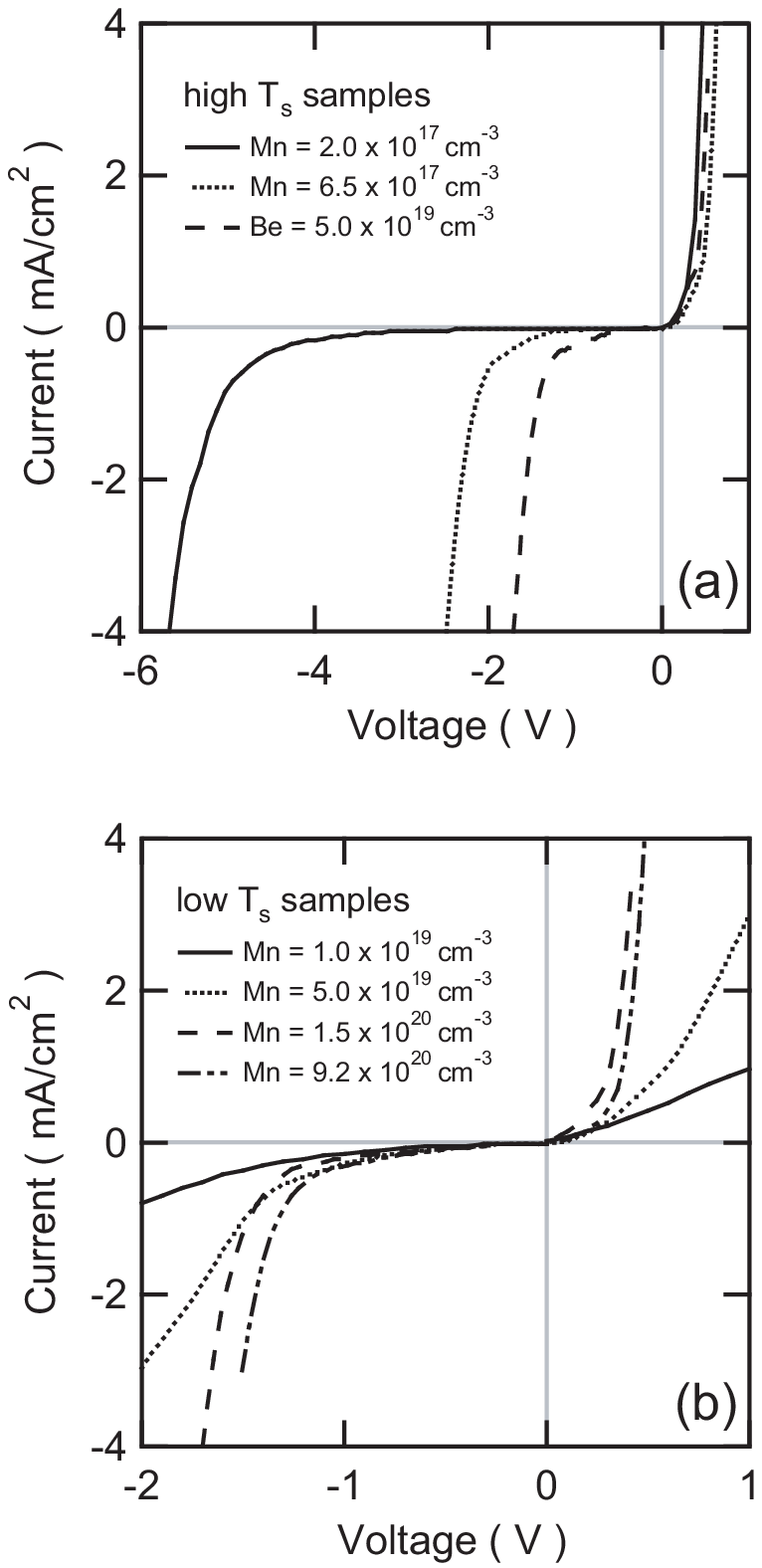}
\caption{\label{I-V} Current-voltage ($I - V$) characteristics of various ES/GaAs:X (X=Mn and Be) junctions at
room temperature. GaAs:X epitaxial layers were grown by molecular beam epitaxy at substrate temperatures of (a) $T_s$
=580 ‹C (high $T_s$ samples) and (b) $T_s$ = 250 ‹C (low $T_s$ samples).}
\end{figure}

We first show current-voltage ($I - V$) and capacitance-voltage ($C - V$) characteristics of the junction composed
of $p$-GaAs and electrolyte solution. $I - V$ curves in the dark are shown respectively in Figs.~\ref{I-V}(a) and (b) for samples
grown respectively at high ($T_s$ = 580 ‹C) and low ($T_s$ = 250 ‹C) substrate temperatures with various Mn and Be
concentrations. The value shown in the $x$ axis is the bias voltage applied to a semiconductor with respect to the
reference electrode. It is obvious from both figures that all samples, regardless of dopant species, show rectification
characteristics with positive bias being the polarity for a large current. This fact indicates that junctions can be
treated as $p$-type Schottkey diodes \cite{16} for which positive bias is the forward direction. In particular, for low
$T_s$ samples, the slope of $I - V$ curve in the forward direction increases with increasing the doping concentration,
reflecting the enhanced sample conductivity. In the negative bias region, the tunneling current across the thin depletion
region is well suppressed presumably because of the lack of available empty states in an electrolyte solution (ES).
Nevertheless, the breakdown voltage in the negative bias region gradually decreases with increasing doping concentration.
Extrapolating forward current density to zero-voltage, saturation current can be extracted, from which the zero-field
barrier height at the ES/semiconductor interface can be estimated on the
basis of thermionic emission model.
The barrier heights thus obtained from various samples are centered around 0.62 $\pm$ 0.04 eV. These values are consistent
with those reported for conventional metal/$p$-GaAs Schottky diodes \cite{22}.
\begin{figure}
\includegraphics{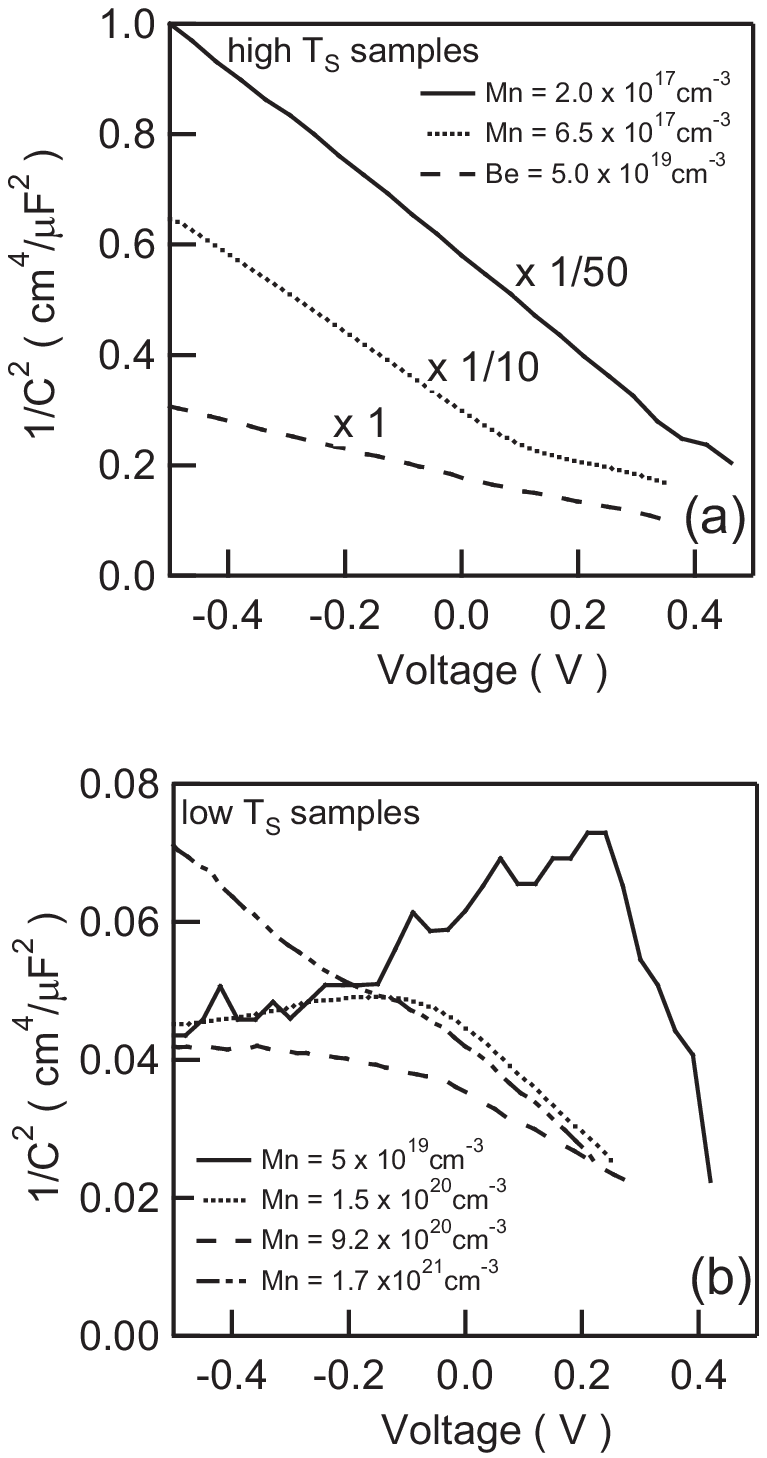}
\caption{\label{C-V}$1/C^2-V$ characteristics for ES/GaAs:X (X = Mn and Be) junctions at room temperature.
Modulation ac-frequency was 5 kHz. GaAs:X epitaxial layers were grown by molecular beam epitaxy at substrate
temperatures of (a) $T_s$ = 580 ‹C (high $T_s$ samples) and (b) $T_s$ = 250 ‹C (low $T_s$ samples).}
\end{figure}

Figs.~\ref{C-V}(a) shows$1/C^2 - V$ relationships for $p$-type, high $T_s$ samples ($T_s$ = 580 ‹C). Linear relationship isclearly
seen for wide range of bias voltage. There is almost no $\omega$ dependence throughout the entire frequency range
($1 - 25$ kHz) used in this work. A slope decreases monotonously with increasing doping concentration irrespective of
dopant species. The number of ionized Mn, $N_{Mn^-}$, estimated from eq.(1) coincides well with the incorporated Mn
concentration $N_{Mn}$. The barrier height extracted from the intercept with the $x$ axis is 0.4 - 0.6 eV, being
consistent with values obtained from $I$-$V$ data. These facts indicate that a well defined depletion region is
established at an ES/semiconductor junction for samples grown at high $T_s$.

On the other hand, for low $T_s$ samples ($T_s$ = 250 ‹C), a straight $1/C^2 - V$ relationship can be established for
a limited bias region defined by the bias voltage of - 0.1 V or higher (Fig.~\ref{C-V}(b)). As to the negative bias region
lower than - 0.1 V, the relationship becomes strongly nonlinear, making it difficult to analyze by the Schottkey plot.
In an extreme case, as seen for the sample with Mn concentration of 5 $\times$ 10$^{19}$ cm$^{-3}$, the straight relationship
appears just in between 0.3 and 0.4 V. A strong deviation from the linear relationship in the negative bias region,
which also occurs for GaAs:Be samples grown at low $T_s$, is probably due to relatively deep interface/bulk states
associated with the low temperature growth. In fact, the nonlinear region is strongly $\omega$ dependent ($\omega$ = $10 - 25$ kHz),
whereas the linear region is rather independent of $\omega$. $N_{Mn^-}$ deduced from the linear region matches well with $N_{Mn}$
(Fig.~\ref{NandN-}). Moreover, the barrier heights extracted from the linear region are around $0.5 - 0.6$ eV, being consistent with
values extracted from $p$-type samples grown at high $T_s$. On the basis of these findings, we conclude that $N_{Mn^-}$ for
low $T_s$ samples can be extracted by carefully analyzing the $C - V$ data. We confirm that the results disclosed
recently in rapid communication \cite{12} are convincing.

	\subsection{Relation between $N_{Mn}$ and $N_{Mn^-}$}
	\label{3-2}

\begin{figure}
\includegraphics{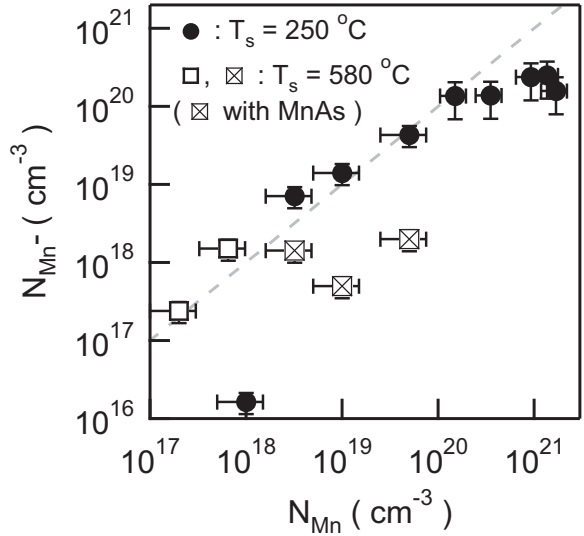}
\caption{\label{NandN-}Relation between incorporated Mn concentration $N_{Mn}$ and ionized Mn acceptor concentration
$N_{Mn^-}$. Dashed line indicates the relation $N_{Mn}$ = $N_{Mn^-}$.   and œrepresent $N_{Mn^-}$ values for high and
low $T_s$ samples, respectively. $\boxtimes$ represents $N_{Mn^-}$ values for high $T_s$ sample containing MnAs.}
\end{figure}

The relation between $N_{Mn}$ and $N_{Mn^-}$ is summarized in Fig.~\ref{NandN-}. Open squares and black circles represent $N_{Mn^-}$
values for high and low $T_s$ samples, respectively. A dashed line in the figure depicts the relation $N_{Mn^-}$ = $N_{Mn}$.
For high $T_s$ samples, $N_{Mn^-}$ increases monotonously with $N_{Mn}$ up to around the mid range of 10$^{18}$ cm$^{-3}$ with
the relation $N_{Mn^-}\approx N_{Mn}$. This fact indicates that incorporated Mn atoms become accepters and are fully
ionized in the depletion region, being consistent with the works done in 70's. For $N_{Mn}$ higher than mid 10$^{18}$ cm$^{-3}$,
$N_{Mn^-}$ does not increase any further but shows saturated behavior. Magnetization measurements for samples in this
region revealed the formation of ferromagnetic MnAs second phase \cite{23}. From these observations, we conclude that
saturated $N_{Mn^-}$ value in $N_{Mn}$ $\geq$ 3 $\times$ 10$^{18}$ (squares with crosses in Fig.~\ref{NandN-}) is due to the
precipitation of MnAs.

As to the low $T_s$ (Ga,Mn)As samples, the relation between $N_{Mn}$ and $N_{Mn^-}$ can be classified into three
different regions. The first region is defined by $N_{Mn}$ $\leq$ 10$^{18}$ cm$^{-3}$ in which the $N_{Mn^-}$ value
is significantly lower than the $N_{Mn}$ value. In this region, incorporated Mn acceptors are compensated by donors
of some sort. One likely source is excess As incorporated during the epitaxial growth at low $T_s$ \cite{24}.
The second region is defined by $N_{Mn}$ = mid 10$^{18} - 10^{20}$ cm$^{-3}$ in which $N_{Mn^-}$ increases
with $N_{Mn}$ with the relation $N_{Mn^-}\approx N_{Mn}$. This is consistent with the presently accepted picture \cite{5,6}
in that Mn atoms are incorporated substitutionally in the group III sub-lattice sites. As to the carrier transport,
however, we infer that impurity conduction is the dominant mechanism in this region, rather than the conduction due to
delocalized holes. This point will be discussed in the next section. The third region is defined by $N_{Mn}\geq 2\times 10^{20}$ cm$^{-3}$
in which $N_{Mn^-}$ begins to saturate. Within the limit of this study, ferromagnetic MnAs second phase was not detected
in low $T_s$ samples even with the highest $N_{Mn}$ of 1.5 $\times$ 10$^{21}$ cm$^{-3}$.
This fact indicates that the saturation behavior in the third region is not due to the MnAs precipitates, but is attributed
to other microscopic origins, such as interstitial Mn, excess As incorporation \cite{2,25}, and Mn-Mn dimers \cite{26}.
The first and second phenomena are believed to result in the formation of deep donors whereas the third one might result
in inert acceptor complex. Very recently, the presence of interstitial Mn has been detected by the Rutherford
backscattering experiments \cite{27}.

It should be noted that the samples in the third region exhibit ferromagnetic order at low temperatures. Curie
temperatures $T_c$ of those samples ($T_c$ = $30 - 60$ K), however, do not follow the empirical formula $T_c$ [K]
= 2000$\times x(=9\times 10^{-20}\times N_{Mn})$ established in ref.25, but rather show a weak $N_{Mn}$
dependence. When $N_{Mn^-}$ is used instead of $N_{Mn}$, we find that experimental $T_c$ matches with this formula.
For samples with higher $T_c$, we notice that discrepancy between $N_{Mn}$ and $N_{Mn^-}$ becomes smaller.
These facts indicate that the discrepancy appearing in the third region is sample dependent and can be interpreted as
the quality of samples.

	\subsection{Relation between $N_{Mn^-}$ and $p$}
	\label{3-3}

\begin{figure}
\includegraphics{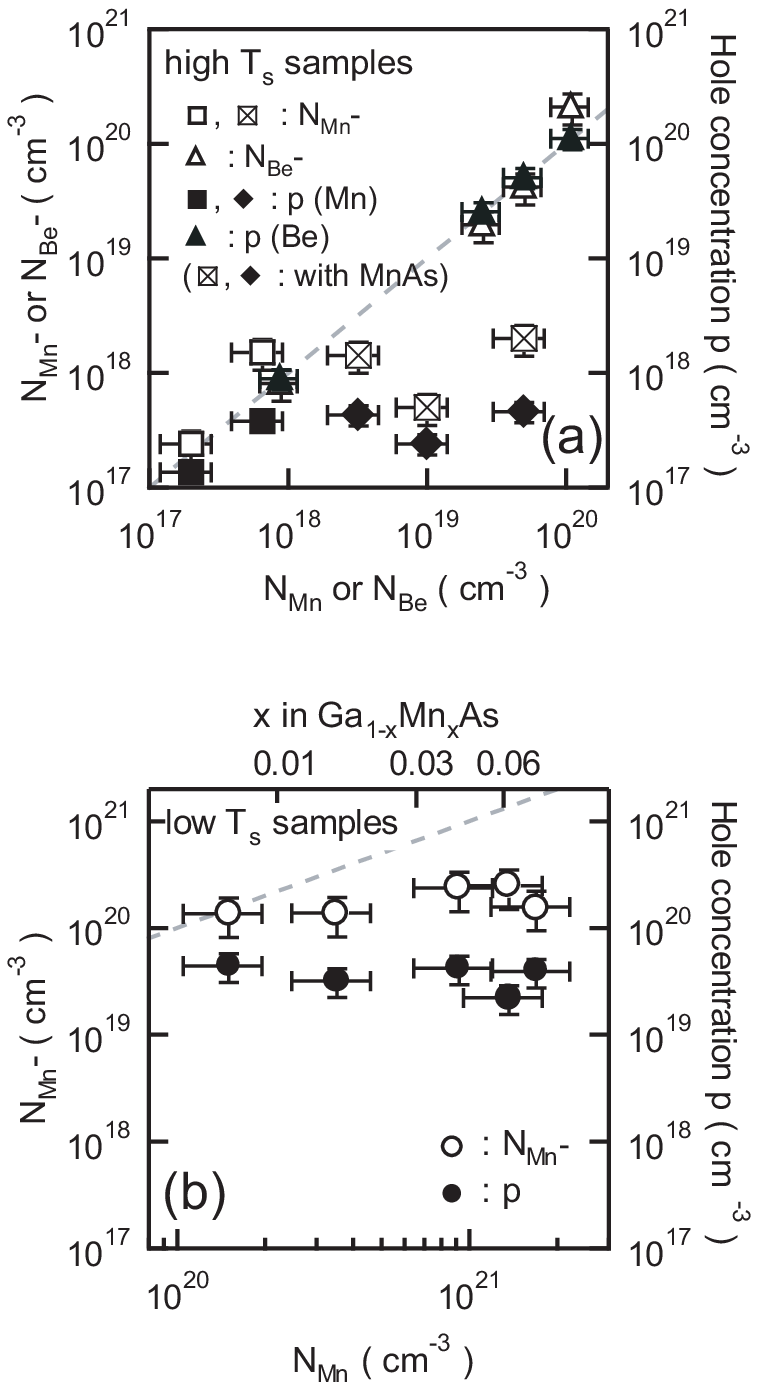}
\caption{\label{Nandp}$N_{Mn^-}$ ($N_{Be^-}$) and $p$ shown as a function of $N_{Mn}$ ($N_{Be}$) for samples
grown at (a) $T_s$ = 580 ‹C (high $T_s$ samples) and (b) $T_s$ = 250 ‹C (low $T_s$ samples). In Fig.4(a),   and ¡
respectively represent $N_{Mn^-}$ and $p$ values for homogeneous samples, whereas $\boxtimes$ and Ÿ represent $N_{Mn^-}$
and $p$ for samples containing MnAs, respectively. In Fig.4(b),› and œ respectively depict $N_{Mn^-}$ and $p$ for
magnetic (Ga,Mn)As samples. The data obtained for $p$-GaAs:Be samples, as represented by ¢ and £, are also plotted
for comparison.}
\end{figure}

\begin{table*}
\caption{\label{table1}Summary of $N_{Mn}$ ($N_{Be}$), $N_{Mn^-}$ ($N_{Be^-}$), resistivity $\rho $,
hole concentration $p$, and mobility $\mu_p$, together with sample ID number, substrate temperature $T_s$ and
Curie temperature $T_c$. It is interesting to note that actual mobility would become three to five times lower than
the Hall mobility if we use $N_{Mn^-}$ as actual $p$ values for magnetic samples (MM142 and M66). }
\begin{ruledtabular}
\begin{tabular}{cccccccc}
Sample No.&$T_s$ (‹C)&$N_{Mn}$ or $N_{Be}$ (cm$^{-3}$)&$N_{Mn^-}$ or $N_{Be^-}$  (cm$^{-3}$)&$T_c$ (K)&
$\rho$ ($\Omega\cdot$cm)&$p$ (cm$^{-3}$)&$\mu_p$ (cm$^2$/V$\cdot$s) \\ \hline
\\
 MM140 (Mn)&580&2.0 $\times$ 10$^{17}$&2.4 $\times$ 10$^{17}$&-&0.18&1.4 $\times$ 10$^{17}$&259\\
 MM153 (Mn)&580&5.0 $\times$ 10$^{19}$&2.0 $\times$ 10$^{18}$&-&0.058&4.6 $\times$ 10$^{17}$&231\\
 M105 (Be)&580&5.0 $\times$ 10$^{19}$&4.2 $\times$ 10$^{19}$&-&0.0017&5.0 $\times$ 10$^{19}$&70.1\\
 MM142 (Mn)&250&1.5 $\times$ 10$^{20}$&1.4 $\times$ 10$^{20}$&30&0.019&4.4 $\times$ 10$^{19}$&7.31\\
 M66 (Mn)&250&9.2 $\times$ 10$^{20}$&2.4 $\times$ 10$^{20}$&55&0.011&4.2 $\times$ 10$^{19}$&14.0\\
\end{tabular}
\end{ruledtabular}
\end{table*}
We now discuss the relationship between $N_{Mn^-}$ and hole concentration $p$. Those quantities are plotted together
as a function of $N_{Mn}$ in Figs.~\ref{Nandp}(a) and (b) for high and low $T_s$ samples, respectively. Open and closed squares
respectively represent $N_{Mn^-}$ and $p$ for high $T_s$ samples (Fig.~\ref{Nandp}(a)), whereas open and closed circles
respectively depict $N_{Mn^-}$ and $p$ for low $T_s$ samples (Fig.~\ref{Nandp}(b)). Squares with crosses and closed diamonds
represent $N_{Mn^-}$ and $p$ for high $T_s$ samples containing MnAs, respectively (Fig.~\ref{Nandp}(a)). The data obtained
for $p$-GaAs:Be samples, as represented by open and closed triangles, are also plotted in Fig.~\ref{Nandp}(a) for comparison.
The data obtained from Hall effect measurements for representative samples are summarized in Table 1 together with $T_s$,
$N_{Mn}$, $N_{Mn^-}$, and $T_c$ values. For $p$-GaAs:Be samples (Fig.~\ref{Nandp}(a)), it is clear that $N_{Be^-}$ values coincide
with both $N_{Be}$ and $p$ up to very high doping concentration, reflecting the fully ionized condition in both depletion
and neutral regions because of the shallow acceptor level of Be \cite{28}. On the other hand, for high $T_s$ GaAs:Mn
samples, $p$ is always lower than $N_{Mn^-}$ (Fig.~\ref{Nandp}(a)). This discrepancy can be understood in term of relatively
deep acceptor level of Mn. In the depletion region where Fermi level $E_F$ is far above $E_A$, Mn acceptors are fully
ionized, whereas, in the neutral region where $E_F$ is relatively closer to $E_A$, they are partially ionized. In fact,
putting the experimental data in eq.(~\ref{4a}) and (~\ref{4b}), we obtain $E_A$ = $50 - 60$ meV together with $E_F$ = 108 and 85 meV
above the top of the valence band for samples with $N_{Mn}$ = 2 $\times$ 10$^{17}$ and 6 $\times$ 10$^{17}$ cm$^{-3}$,
respectively. The $E_A$ values estimated in this fashion are consistent with earlier works in which $E_A$ = $60 - 80$ meV
for 10$^{17-18}$ cm$^{-3}$ \cite{8,9}. This fact indicates that anomalous Hall effect is negligibly small in high $T_s$,
Mn-doped samples. As to the $p$-type samples containing MnAs, the relation between $N_{Mn^-}$ and $p$ can also be
understood in terms of Fermi distribution eq.(~\ref{4a}). Influence of MnAs on electronic property is not significant in these
samples.
Let us now turn into (Ga,Mn)As samples prepared at low $T_s$. In view of magneto-transport characteristics, samples are
again classified into three different regions. The first region is defined by $N_{Mn}\leq 10^{18}$ cm$^{-3}$
in which samples are highly resistive, making it difficult to extract $p$ from Hall effect. The second region lies
between $N_{Mn}$ = mid 10$^{18}$ and 10$^{20}$ cm$^{-3}$ in which samples are conductive but show virtually zero Hall
resistance at room temperature (RT) as well as at 77 K. In conventional high $T_s$ GaAs:Mn samples with $N_{Mn}\approx 10^{19}$ cm$^{-3}$,
similar behavior has been observed at low temperatures ($<$ 75 K), and has been discussed in terms of impurity conduction
\cite{9}. Knowing that, for high $T_s$ samples, MnAs precipitation occurs at around $3 \times 10^{18}$ cm$^{-3}$
beyond which $N_{Mn^-}$ saturates (Fig.~\ref{NandN-}), the effective $N_{Mn}$ that causes the impurity conduction is inferred to be
slightly above 3 $\times 10^{18}$ cm$^{-3}$. It is also inferred that, if effective $N_{Mn^-}$ is increased, the
conductivity due to impurity conduction increases so that the temperature that causes a cross-over from impurity
conduction to delocalized-band conduction shifts toward high temperatures. We believe that this is what has been
observed for low $T_s$ samples with $N_{Mn}$ = mid 10$^{18}-10^{20}$ cm$^{-3}$. Consequently, the second region can be
treated as an extended region beyond $N_{Mn}$ = mid 10$^{18}$ cm$^{-3}$ of high $T_s$ samples with $N_{Mn}$ = $N_{Mn^-}$ = $p$.
For low $T_s$ samples, impurity conduction dominates even at RT, reflecting the very high $N_{Mn}$ values.

The third region is defined by $N_{Mn}\geq 2\times 10^{20}$ cm$^{-3}$ in which samples are sufficiently conductive and
exhibit finite, positive Hall resistance. Results are shown in Fig.~\ref{Nandp}(b) and Table~\ref{table1}. As discussed in the previous
paragraph, impurity band is formed $N_{Mn}\sim$ mid 10$^{18}$ cm$^{-3}$ or higher. When $N_{Mn}$ is increased further,
the overlap of wavefunction becomes more and more significant, causing an increase in impurity band width. At some point,
the impurity band overlaps with top of the valence band, resulting in charge transfer between the two bands. Under this
situation, carrier transport occurs through the delocalized valence band mixed with the impurity band and thus revival of
the Hall effect. Based on our experimental results, the overlap between the two bands seems to occur at around
$N_{Mn}\sim 2\times 10^{20}$ cm$^{-3}$. Within the limit of this scenario, it is reasonable to assume that the number of
carriers $p$ is equal to $N_{Mn^-}$, being similar to the second region. However, experimentally, we notice that $p$
obtained from Hall effect is lower than $N_{Mn^-}$ in the third region ($N_{Mn}>N_{Mn^-}>p$). Finally, we come to
the point that the anomalous Hall effect is likely to give rise to the discrepancy between $p$ and $N_{Mn^-}$ in this region.

	\subsection{Asymmetric scattering constant $c$}

\begin{figure}
\includegraphics{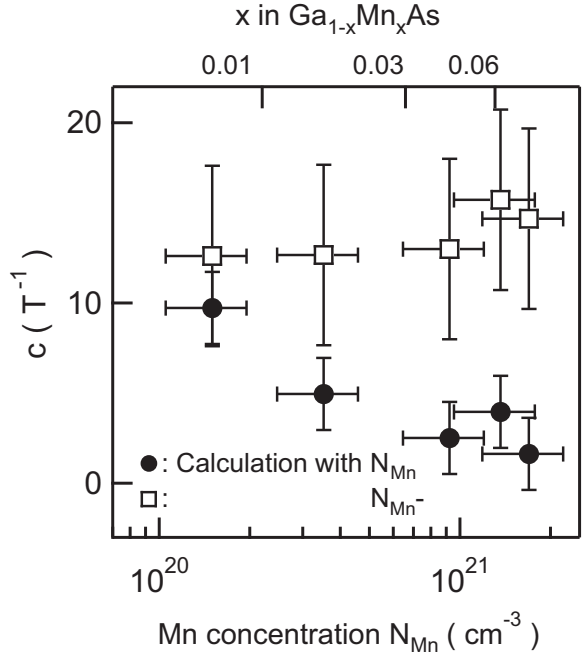}
\caption{\label{asymmetric} A plot of asymmetric scattering constant $c$ vs. incorporated Mn concentration $N_{Mn}$.}
\end{figure}
To pursue further, we have calculated the asymmetric scattering constant $c$ by using eqs.(~\ref{RH}), (~\ref{R0}) and (~\ref{Rs}).
This is the estimation of $c$ at room temperature for the first time. The experimental value obtained from the Hall
effect measurements was substituted directly for $R_H$ in the left hand side of eq.(~\ref{RH}). The first term in the right
hand side of eq.(~\ref{RH}) was estimated from eq.(~\ref{R0}) by substituting experimental $N_{Mn^-}$ value obtained from $C - V$
measurements for $p$. The anomalous Hall term can then be obtained, and it is substituted for the left hand side of
the eq.(~\ref{Rs}). In eq.(~\ref{Rs}), perpendicular magnetization $M$ at room temperature is estimated from the calculated magnetic
susceptibility $c$ based on the Curie-Weiss law ${\chi}= C / (T - \theta _p)$ with $C=N_{Mn}g^2\mu_B^2 \{S (S + 1)\}/3k$.
Here, paramagnetic Curie temperature $\theta_p$ is assumed to be equal to the Curie temperature, whereas spin number $S$
and $g$ factor are $S$ = 5/2 and $g$ = 2, respectively. $n$ = 1 in eq.(~\ref{Rs}) because skew scattering is believed to be
dominant in (Ga,Mn)As \cite{3}. It is very important to notice that the constant $c$ extracted in this fashion from
eq.(3b) can be influenced significantly by the definition of Mn atoms that is responsible for $c$ at RT.
There are two possibilities to estimate $c$; either using the total concentration $N_{Mn}$ or using the concentration
of ionized Mn acceptors $N_{Mn^-}$. The $c$ values calculated for both cases are shown in Fig.~\ref{asymmetric} , from which we notice
an interesting difference between the two cases. When $N_{Mn}$ is used, we obtain $c$ values that range between 1 and
10 and are nearly inversely proportional to $N_{Mn}$. In contrast, in case of $N_{Mn^-}$, we find a fixed value of
$c$ = 13 $\pm$ 5 independent of the $N_{Mn^-}$. Since the range of Mn concentration is low enough to neglect the
scattering processes of higher orders, we believe that $c$ should in principle be independent of $N_{Mn^-}$
(and $N_{Mn}$ as well). Consequently, the results based on $N_{Mn^-}$ seem to grab the essential physics in magnetic
(Ga,Mn)As. Previously, the asymmetric scattering constant was estimated to be $c$ = 2 for Ga$_{\textrm{1-}x}$Mn$_{x}$As
with $x$ = 0.035 by comparing magnetization and Hall resistance data in ferromagnetic phase \cite{3}.
In ferromagnetic $p$-(In,Mn)As, the $c$ value has been estimated to be $c$ = 6 for $x$ = 0.013 \cite{29}.
That the $c$ value extracted in this work is larger than the values extracted in previous works may come from the use
of the $N_{Mn^-}$ instead of $N_{Mn}$. This point should be examined in the future work by carefully evaluating
magnetization data and transport data for wide range of temperature.

\section{Conclusions}

Determination of $N_{Mn}$, $N_{Mn^-}$, and $p$ has been carried out for Mn-doped GaAs epitaxial layers with wide
range of Mn concentrations ($10^{17}-10^{21}$ cm$^{-3}$). Key experiment in this study was electrochemical $C-V$ method
by which $N_{Mn^-}$ has been extracted successfully. Carefully comparing $N_{Mn^-}$ with $N_{Mn}$ and $p$, several
important conclusions have been deduced. For high $T_s$ samples, Mn starts to precipitate in the form of MnAs when
$N_{Mn}$ becomes higher than about $3\times 10^{18}$cm$^{-3}$. Except this point, electronic behavior of the high
$T_s$ epilayers is not affected by the MnAs inclusions, and is essentially the same as those studied in 70's. For
low $T_s$ samples, electronic behaviors can be classified into three different regions. The first region is highly
resistive, strongly compensated region in $N_{Mn}\leq 10^{18}$ cm$^{-3}$. The second region is impurity conduction
region in $N_{Mn}$=mid $10^{18}-10^{20}$ cm$^{-3}$ with $N_{Mn}\approx N_{Mn^-}$, which can be regarded as the extended
region of high $T_s$ epilayers of $N_{Mn}$ beyond mid 10$^{18}$ cm$^{-3}$. The third region, defined by $N_{Mn}\geq 2\times 10^{20}$cm$^{-3}$,
is the region having characteristics similar to the electrical conduction due to delocalized holes. In this region,
magnetism and carrier transport are strongly correlated to each other, causing ferromagnetism and anomalous Hall effect.
Quantitative assessment of anomalous Hall effect at room temperature has been carried out, from which asymmetric
scattering constant $c$ is determined to be $c$ = 13 $\pm$ 5 at room temperature.

\begin{acknowledgments}
We gratefully acknowledge S. Kikuchi for technical assistance. This work is supported in part by Scientific Research in
Priority Areas "Semiconductor Nanospintronics" of The Ministry of Education, Culture, Sports, Science and Technology, Japan. 
\end{acknowledgments}


\begin{thebibliography}{99}
\bibitem{23}
a hyseretic $M - H$ curve with saturation magnetization of 3 $\times$ 10$^{-7}$ emu was detected in a high $T_s$ GaAs:Mn sample with $N_{Mn} = 3 \times 10^{18}$ cm$^{-3}$. About 40 \% of Mn is incorporated in the form of MnAs second phase.
\end{thebibliography}
\end{document}